\documentclass[sigconf,screen,nonacm]{acmart}

\AtBeginDocument{%
  \providecommand\BibTeX{{%
    \normalfont B\kern-0.5em{\scshape i\kern-0.25em b}\kern-0.8em\TeX}}}

\usepackage{subfigure}
\usepackage{multirow}
\usepackage{xcolor}

\usepackage{amsmath}
\usepackage{multicol}
\usepackage{listings}

\definecolor{codegreen}{rgb}{0,0.6,0}
\definecolor{codegray}{rgb}{0.5,0.5,0.5}
\definecolor{codepurple}{rgb}{0.58,0,0.82}
\definecolor{backcolour}{rgb}{0.95,0.95,0.92}

\title[The Elements Project]%
      {Project Elements: A computational entity-component-system in a scene-graph pythonic framework, for a neural, geometric computer graphics curriculum}

\begin{document}

\author{George Papagiannakis}
\orcid{0000-0002-2977-9850}
\affiliation{%
  \institution{FORTH - ICS, University of Crete, ORamaVR}
  \country{}
}

\author{Manos Kamarianakis}
\authornote{Corresponding Author, \url{kamarianakis@uoc.gr}}
\orcid{0000-0001-6577-0354}
\affiliation{%
  \institution{FORTH - ICS, University of Crete, ORamaVR}
  \country{}
}

\author{Antonis Protopsaltis}
\orcid{0000-0002-5670-1151}
\affiliation{%
  \institution{University of Western Macedonia, ORamaVR}
  \country{}
}

\author{Dimitris Angelis}
\orcid{0000-0003-2751-7790}
\affiliation{%
  \institution{FORTH - ICS, University of Crete, ORamaVR}
  \country{}
}

\author{Paul Zikas}
\orcid{0000-0003-2422-1169}
\affiliation{%
  \institution{University of Geneva, ORamaVR}
  \country{}
}

\renewcommand{\shortauthors}{Papagiannakis, Kamarianakis, Protopsaltis et al.}

\begin{abstract}
We present the Elements project, a lightweight, open-source, computational science and computer graphics (CG) framework, tailored for educational needs, that offers, for the first time, the advantages of an Entity-Component-System (ECS)
along with the rapid prototyping convenience of a Scenegraph-based pythonic framework. This novelty allows advances in the teaching of CG: from heterogeneous directed acyclic graphs and depth-first traversals, to animation, skinning, geometric algebra and shader-based components rendered via unique systems all the way to their representation as graph neural networks for 3D scientific visualization. Taking advantage of the unique 
ECS in a a Scenegraph underlying system, this project aims to bridge CG curricula and modern game engines (MGEs), that are based on the same approach but often present these notions in a black-box approach. It is designed to actively utilize software design patterns, under an extensible open-source approach. Although Elements 
provides a modern (i.e., shader-based as opposed to fixed-function OpenGL), simple to program approach with Jupyter notebooks and unit-tests, its CG pipeline is not black-box, 
exposing for teaching for the first time unique challenging scientific, visual and neural computing concepts.
\end{abstract} 

\maketitle



\section{Introduction}

Computer Graphics is a challenging teaching subject \cite{suselo2017journey,Mashxura.2023}, as it borrows knowledge from multiple 
computational scientific areas \cite{balreira2017we} such as mathematics, biology, physics, computer science and specifically software design, data structures and
 GPU programming through shader-based APIs. 
Along with the multiple pedagogical approaches, 
there exist a variety of tools and frameworks \cite{toisoul2017accessible, andujar2018gl, miller2014using, Suselo2019, brenderer,SousaSantos2021,CodeRunnerGL,Wuensche2022,
toisoul2017accessible} that facilitate CG development;
however, only a small subset is oriented towards teaching a hands-on, programming and assignment-based approach of a part 
of or the complete modern 
graphics pipeline, from shader-based visualisation all the way towards neural computing. An approach that seems to yield positive feedback 
from the students is the use of notebooks \cite{SousaSantos2021}, usually based on WebGL, 
that help visualize concepts such as lighting, shadows and textures 
in an interactive way \cite{CodeRunnerGL}. Such tools offer limited functionality as they focus on specific CG aspects, detached from  the CG pipeline 
\cite{Wuensche2022} such as the GLSL shaders \cite{toisoul2017accessible}, or 
raytracing \cite{vitsas2020rayground}. 
In such approaches, students may focus and learn in 
depth how a particular task is performed or how specific parameters 
affect the final rendered scene, but they usually have trouble comprehending  
how these operations are interconnected into a single pipeline. 
Another strategy to teach the entire operation of the OpenGL 
pipeline is via the employment of frameworks \cite{andujar2018gl, miller2014using, brenderer}. These packages 
usually contain examples and helper 
functions that perform batch-calling of simpler OpenGL functions, 
based on their functionality, e.g., shader initialization or 
VAO/VBO creation, 
thus facilitating students understanding without exposing them to 
low-level code, until they are ready to handle or tamper with it. 
Such frameworks are usually written in C++/C\# and are suitable both 
for teaching as well as small or intermediate projects. A main drawback of them is that 
several CG pipeline operations are treated as black-boxes, thus not providing the practical knowledge of all stages of the rendering pipeline to students, like the calculation of the global transformation matrix of a 3D mesh in a scenegraph. Furthermore, these frameworks are not 
extensible
in a simple, straightforward and open-source way towards state-of-the-art python-based neural computing for geometric deep learning frameworks such as graph neural networks.

\subsection{A rising tide: ECS and Scenegraphs}

The management of interactive scenes in MGEs and 
graphics systems is based on a hierarchical data structure, the 
\emph{scenegraph}, a heterogeneous, directed acyclic graph, that is traversed to render a 
frame \cite{rohlf1994iris}. The scenegraph contains all the data required to replicate the 
scene, including among others, geometry, materials, camera 
specifications and light information. All graph nodes inherit 
properties from their ancestors and the object mesh data of a specific 
node are stored on child leaf nodes. The nodes are usually referred to 
as \emph{gameobjects}, \emph{actors} or plain \emph{objects} in MGEs, 
whereas the related data stored within them are usually denoted 
as \emph{Components}. Scenegraph edges denote connectivity and hierarchy;
different traversals ensure initialisation, update, culling and rendering of all scenegraph nodes.

Entity-Component-System is an architectural pattern, mainly used in 
3D applications and game development \cite{Nystrom2014,graph2}, that 
decouples data from behavior, simplifying the development. It heavily 
relies on data-oriented-design and composition, where created entities 
are assigned independent components, in contrast with object-oriented-
design, where  such components would be inherited from base classes.
This architecture enables a) improved performance in applications with 
many objects (e.g physics based simulations) and b) improved 
maintenance and understanding of the application's objects.

The ECS model has gained a lot of attention in various game 
frameworks (EnTT, \url{https://github.com/skypjack/entt}), \cite{Polyphony} and a variation of it is in the core of most MGEs. 
For example, Unity is currently restructuring its core-engine, towards adopting the DOTS (acronym for Data Oriented 
Technology Stack) system (\url{https://unity.com/dots}) that features an ECS architecture, to tackle 
violations caused by previous data-oriented programming principles and achieve better fps performance in complex scenes. In our work, 
we provide a unique core ECS in a scenegraph architecture, in order to teach several key CG concepts in a data-oriented approach that facilitates connectivity with neural computing and geometric deep learning.

\subsection{Current approaches in modern CG teaching} 
\label{sub:motivation}

To enable a CG hands-on programming approach, students are usually given a 
framework  and initial pipeline skeleton examples, upon which they may build 
up their first CG applications. Such approaches utilize a MGE, or a 
C/C++ framework.

MGEs, such as Unity and Unreal engine, are high-level tools 
that provide excellent visualizations of content using all CG 
principles, but most low-level details of the CG pipeline are 
kept hidden in a black-box. For instance, any Unity programmer 
may easily create CG content, such as a shadowed cube with no 
knowledge in lighting or shadow algorithms, or animate a simple 
model without having to deal with the frame (linear or spherical) 
interpolation process and the calculation of the world 
transformation matrix, or rotate an object without knowing the 
details of Euler angles or quaternions. In that frame, as the 
learning objectives of all CG courses include the low-level 
experimentation of students with all basic CG principles, 
CG professors have suspended the use of game engines in 
undergraduate CG courses. On the other hand, game engines 
are widely used in graduate CG courses, where students have 
already mastered CG principles well.

In many CG curricula, professors have opted 
for the use of C/C++ CG frameworks or packages, prioritizing the learning
of basic principles of the  CG pipeline (rendering, lighting, shading).
Although frameworks are often hierarchically structured and code exploration is, in some cases, possible at some levels, the complexity of such frameworks is great, hindering swift prototyping and connection with rapidly growing scientific domains. Furthermore, a fully educational-tailored experience at all levels is still not feasible, and state-of-the-art concepts, such as ECS on a scenegraph and design patterns are not utilized yet.


\subsection{Our pythonic, ECS in a Scenegraph, approach: ECSS} 
\label{sub:overview_our_approach}

In this paper, we introduce the \emph{Elements} project, that aims to 
provide the power of the Entity-Component-System (ECS) in 
a Scenegraph (ECSS) \emph{pythonic} (i.e., using latest python 3 language features and software design patterns) framework, suitable to be used in the 
context of a CG curriculum with extensions to scientific, neural computing and geometric deep learning. 

As a central idea, we have implemented a software-design-pattern based scenegraph, combined with an Entity Component System (ECS) \cite{bilas2002data}
principle, which dictates a data-oriented rather than an object-oriented 
programming approach. By opting to ECS, we were able to serve the main 
approach that MGEs abide to and 
grants them the ability to effectively manage even extremely 
complicated scenes. 

In our ECSS, the functionality of our scenegraph is provided via the Composite pattern, our Entities are just Grouping nodes with an ID, our Components contain only data that can be easily extended 
using the decorator pattern, and all functionality (rendering, update etc.) is only performed in our Systems that are traverse the scenegraph of Entities and Components
via the visitor pattern. Finally, events and messages in our ECSS  can easily be handled using the observer pattern. 

\begin{figure}[tbp]
   \centering
   \includegraphics[width=0.47\textwidth]{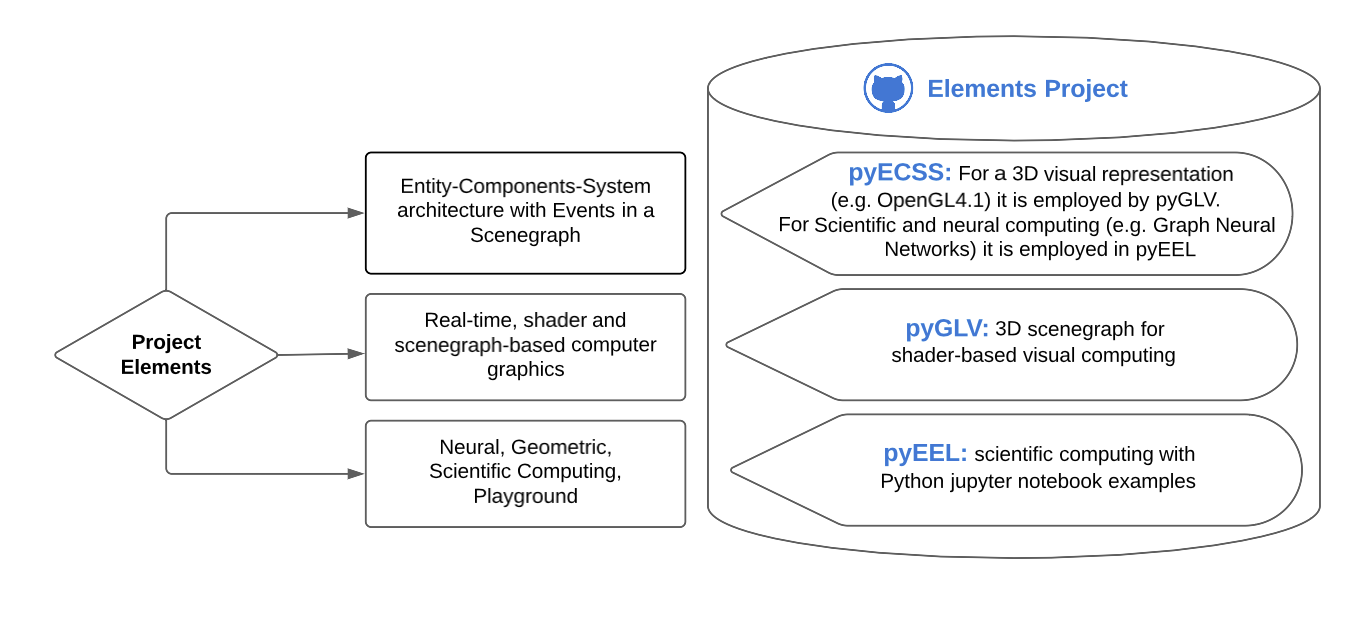}
   \caption{The Elements Project: pyECSS, pyGLV \& pyEEL.}
   \label{fig:Elements1}
\end{figure}

Although scenegraphs, shader-based GPU programming and ECS are well known principles in the CG 
community, the existence of a framework that incorporates all 
of them and exposes them as easily accessed notebooks for scientific and neural computing has not appeared in the bibliography so far, to the best of our knowledge. 
Included in Elements (see Fig. \ref{fig:Elements1}), the pyECSS (acronym for \textbf{py}thon 
\textbf{ECS} on \textbf{S}cenegraphs), is a unit-test python package, that, for the first time, incorporates both of these principles, 
in a comprehensive way, suitable for teaching to CG students. 

The pyGLV (acronym for computer \textbf{G}raphics for deep 
\textbf{L}earning and scientific \textbf{V}isualization in 
\textbf{py}thon) package of Elements exploits the pyECSS 
benefits and provides the necessary graphics algorithms to 
allow the creation and visualization of 3D scenes via the OpenGL 
and GLSL APIs and languages. The use of simple functions allows the creation of 
 Entities (scene root and objects) and their Components (geometry, 
camera, lights, transformation) of objects in the 3D scene, CG programmers can 
quickly dive into the concept of a scenegraph, an approach that will help 
them easily adapt to the pipeline of MGEs. The built-in 
Systems (TrasformationSystem, CameraSystem) and the shader functions prove 
to be essential for beginners, as they help them grasp the CG principles 
without initially worrying about low-level implementation details. 

Lastly, the Elements' pyEEL (acronym for \textbf{E}xplore - \textbf{E}xperiment - 
\textbf{L}earn using \textbf{py}thon) repository holds various 
jupyter notebooks, suitable for both beginners and experienced programmers. 
Our aim is to evolve this repository to a knowledge-hub, a place where 
tutorial-like notes will help the reader comprehend and exploit the pyECSS and/or the pyGLV 
package to their full extent. As many research frameworks in diverse scientific and neural domains are python based, we envision to further contribute in 
such scientific areas, as Elements is best suited for data visualization, immersive analytics and geometric deep learning with graph neural networks (GNNs). 

\subsection{The pyECSS package} 
\begin{figure}[tbp]
   \centering
   \includegraphics[width=0.48\textwidth]{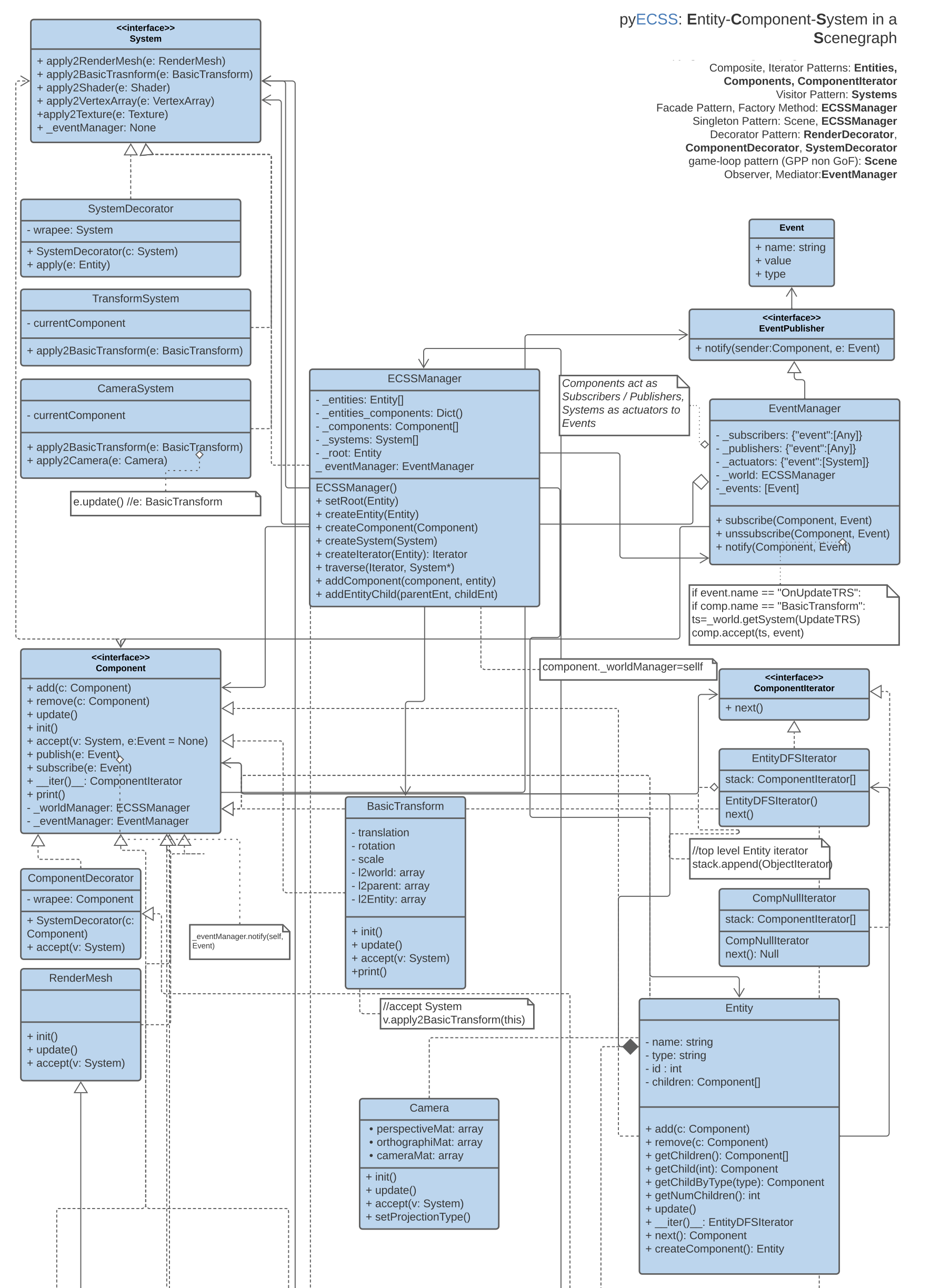}
   \caption{pyECSS class diagram}
   \label{fig:pyECSS}
\end{figure}

\subsection{Benefits of using the Elements projest in CG curricula} 
\label{sub:benefits_of_using_elements}


Below is a list of the benefits from using Elements in a CG class, 
both as an instructor as well as a student.

\begin{enumerate}
   \item \textbf{ECS in a Scenegraph (ECSS).} Teaching the benefits of 
   ECS when applied in a scenegraph will allow the students to 
   understand the main CG concepts using a rendering pipeline 
   that is very similar to the ones utilized by MGEs, which can support rendering million of objects 
   in real-time. This will assist students on organizing and 
   designing a scene or on incorporating new features. 
   As individual Systems can work autonomously and in parallel 
   with other Systems, a faster rendering can be obtained. 
   This ECSS approach can also be followed in other 
   domains besides CG, or for other CG tasks besides scene rendering, like physics computations. To this end, ECSS functionality 
   is encapsulated in the standalone pyECSS python package.
   \item \textbf{A python-based, visual computing framework based exclusively on software design-patterns.} Python is a language that is both ideal for novice and 
   proficient CG programmers as well as becoming a defacto standard in neural computing. Its versatility comes from the rapidly growing variety of developed scientific libraries and frameworks, by its large active community 
   of developers and scientific computing domain experts, that exploit python's rapid prototyping ability. Much of the functionality of the  Elements' packages 
   is simplified via unit-tested methods and classes, requiring minimal initial setup and actions by a novice CG programmer to obtain a 
   completely rendered frame. 
   Despite its simplicity, such functionality is provided 
   through a white-box CG pipeline, where all its intermediate steps may be accessed and tampered. The use of software design patterns  allows to easily extend the features supported by Elements. In that respect, the implementation of  a new component and the respective system that traverses it, is now a simple task due to the use of ECS; the resulting extensibility potential makes the project future-proof. 
   \item \textbf{Versatile CG teaching approaches}:  
   The Elements' framework allows both typical bottom-up approaches, 
   explaining the basic function 
   and tools one by one and then assembling them together, as well as also 
   top-to-bottom approaches, thus starting from the creation 
   of a scene-graph and its visualization and sequentially explain the 
   under-the-hood CG pipeline algorithms and operations. 
   \item \textbf{Tailored for Education}: Elements is a lightweight system, 
   tailored for training needs and available for free as open-source. As opposed 
   to the use of commercial MGEs, which are proprietary products of huge magnitude, completeness and complexity, the proposed project offers a more 
   targeted educational tool, suitable for novice and
   intermediate CG programmers. Its design enables students to dive deep 
   into the rendering pipeline in a simple and distraction-less way, that is suitable 
   for an undergraduate CG course.
\end{enumerate}

\section{The key components of the Elements project}

The pyECSS python package, based on pure software design patterns, features 
a plain but powerful CG architecture with Entities, Components and Systems 
in a Scenegraph.
A class diagram of the pyECSS package and their in-between 
dependencies is shown in Fig.~\ref{fig:pyECSS}.
The pyECSS package, after a simple installation via PyPi, allows the programmer to easily create a scenegraph 
using Entities and Components as nodes. A simple example is shown 
in Listing~\ref{lst:sample}, where several entities 
are created and interconnected via parent-child relationships and 
component nodes are added to each entity.

Within the developed ECSS framework entities act as unique 
identifiers, components contain the associated data and 
systems are classes that operate on both of them,
performing various tasks, such as updating the position of an entity based
on the root or passing a visible object to the vertex shader. 

The advantage of using an ECSS is that it allows developers to build 
flexible, modular systems that can easily be extended and modified. 
It also makes it easier to reason about the interactions between different 
objects in the scene, since each object is represented by its own set of 
components. Such ECSS architectures are particularly useful in 
CG and gaming applications, where scenes with large number of objects, each 
with complex behaviours, that need to be updated and rendered in 
real-time, is a common issue.

\begin{figure}[tbp]
   \centering
   \includegraphics[width=0.45\textwidth]{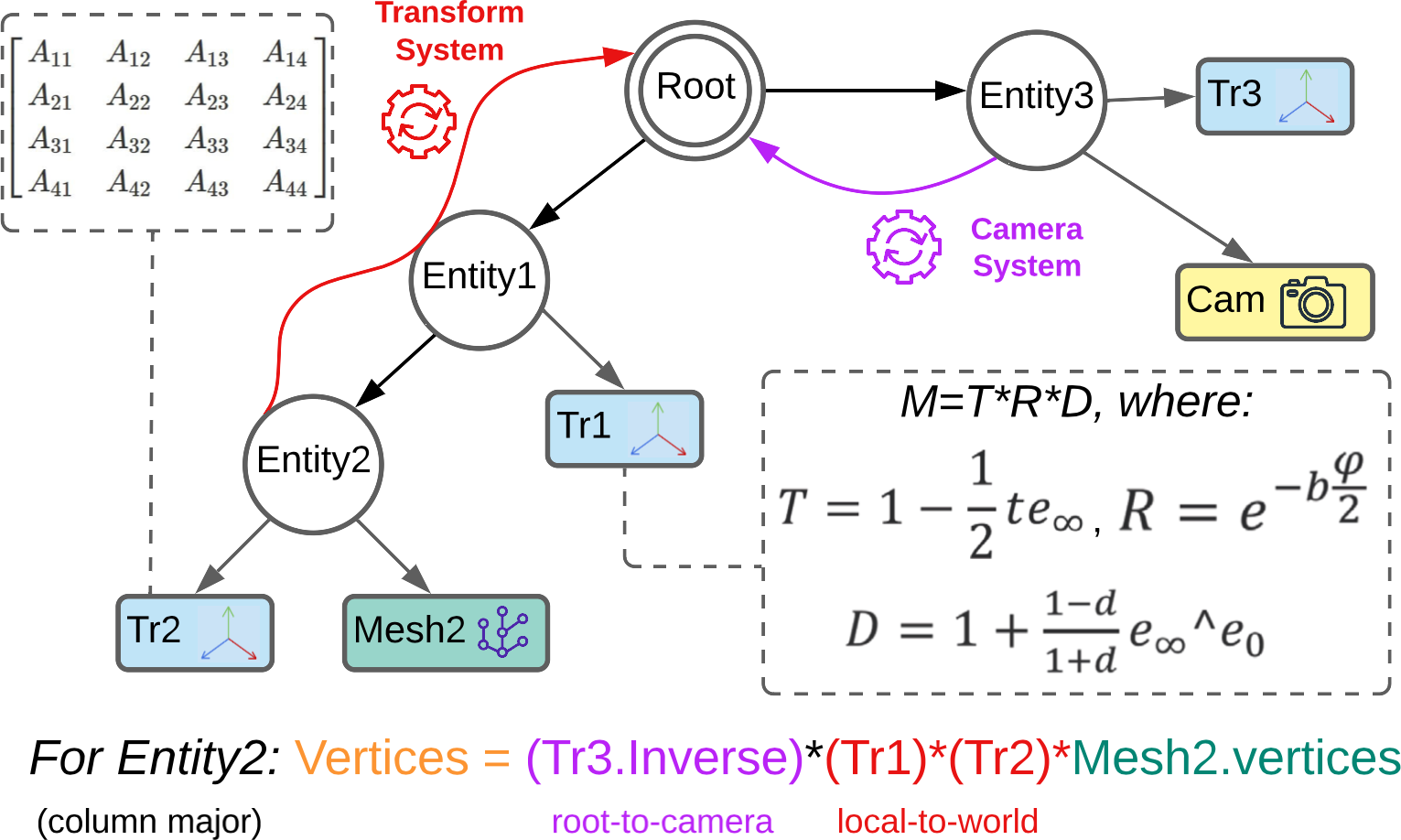}
   \caption{A simple scenegraph based on ECS. Circular nodes 
   correspond to 
   Entities while rectangular correspond to Components. The directed 
   edges imply a parent-to-child relationship. Trasform system evaluates the local-to-world matrix for all entities the root-to-camera is determined via the Camera System. Transformation data
   for Entity1 is provided in matrix form, whereas CGA multivectors 
   are employed for Entity2; $e_\infty, e_o$ are basis 
   multivectors, $t,b$ are the translation vector and the 
   rotation axis expressed as multivectors, $\phi$ 
   is the rotation angle, $d$ is the dilation factor 
   and $e,\wedge$ denote the exponential and wedge product functions.}
   \label{fig:scenegraph}
\end{figure}

\lstdefinestyle{mystyle}{
    backgroundcolor=\color{backcolour},   
    commentstyle=\color{codegreen},
    keywordstyle=\color{magenta},
    numberstyle=\tiny\color{codegray},
    stringstyle=\color{codepurple},
    basicstyle=\footnotesize,
    breakatwhitespace=false,         
    breaklines=true,                 
    captionpos=b,                    
    keepspaces=true,                 
    numbers=left,                    
    numbersep=5pt,                  
    showspaces=false,                
    showstringspaces=false,
    showtabs=false,                  
    tabsize=2
}
\lstset{style= mystyle, language=Python}
\lstset{frame=lines}
\lstset{caption={ Scenegraph creation code snippet where we add and connect entities and components.}}
\lstset{label={lst:sample}}
\lstset{basicstyle=\tiny\ttfamily}
\begin{lstlisting}
# Adding and Connecting Entities
scene = Scene()
sw = scene.world
root = sw.createEntity(Entity(name="Root"))
entity1 = sw.createEntity(Entity(name="Entity1"))
sw.addEntityChild(rootEntity, entity1)
entity2 = sw.createEntity(Entity(name="Entity2"))
sw.addEntityChild(entity1, entity2)
entity3 = sw.createEntity(Entity(name="Entity3"))
sw.addEntityChild(rootEntity, entity3)
...
# Adding Components to Entities
trans1 = sw.addComponent(entity1, BasicTransform())
shader2 = sw.addComponent(entity1, Shader())
vArray2 = sw.addComponent(entity1, VertexArray())
mesh2 = sw.addComponent(entity1, RenderMesh())
orthoCam = sw.addComponent(entity2, mycamera)

\end{lstlisting}

\subsection{The pyGLV package}
\begin{figure}[tbp]
   \centering
   \includegraphics[width=0.48\textwidth]{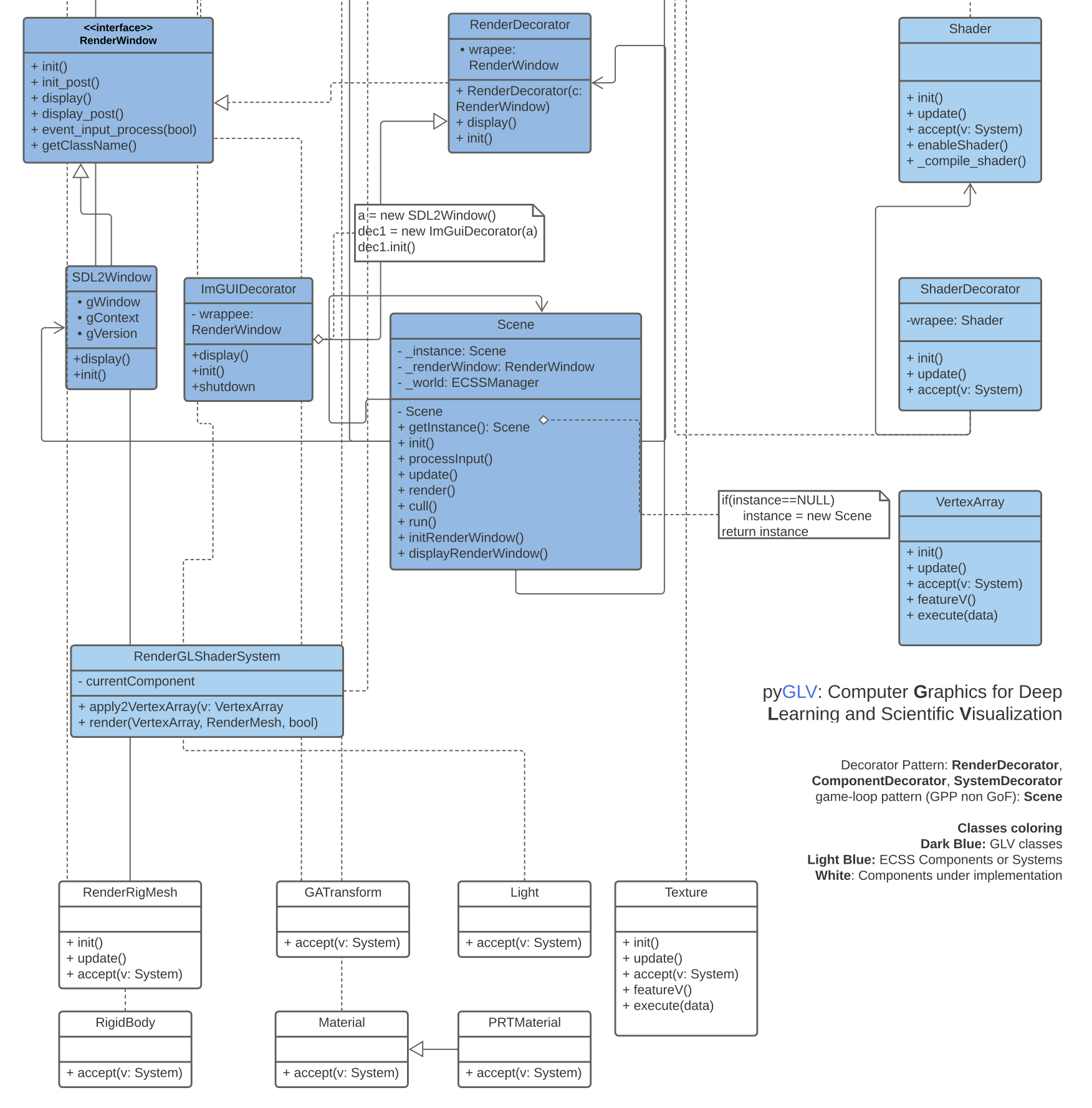}
   \caption{pyGLV class diagram}
   \label{fig:pyGLV}
\end{figure}

The pyGLV python package, based on pure software design patterns, utilizes the previous ECSS architecture, via the pyECSS package. Its main aim is to showcase the applicability of basic, cross-platform, OpenGL-based, real-time computer graphics, in the fields of scientific visualization and geometric deep learning.
A class diagram of the pyGLV package and their 
in-between dependencies is shown in Fig.~\ref{fig:pyGLV}.

The pyGLV contains the implementation of many built-in components,
that are used to setup CG scenegraphs, using the pyECSS package.
\begin{itemize}
\item \textbf{BasicTransform.} A component that stores the objects' 
coordinate system with respect to its parent entity, in the form of a 
product of a translation, rotation and scaling matrix. 
\item \textbf{Camera.} A component storing the camera setup information 
in the form of a view matrix; it can be generated via the \texttt{ortho} or 
\texttt{perspective} functions, contained in the pyECSS.utilities package. 
\item \textbf{RenderMesh.} A component that stores the basic geometry of an entity, such as the vertex positions and face indices of an object.
It may also contain the colour of the object or the colour of each 
vertex, as well as the normals of each face triangle. 
\item \textbf{VertexArray.} A component used to store the vertex array 
object (VAO) and vertex buffer object (VBO) for entities that will be 
passed to the vertex shader.
\item \textbf{Shader.} A component that stores the data required for an 
OpenGL-GLSL Shader; it contains basic vertex and fragment shaders ready 
to be used. 
\end{itemize}

The code segment in Listing~\ref{lst:sample}, depicts the attachment of BasicTransform, RenderMesh, a VertexArray and Camera components  on 
various entities of the scene.
Depending on the situation, new components may be created to store 
different types of data, while existing components may be decorated to 
allow the storage of additional data of the same or equivalent representation  
formats, such as store transformation data in alternative algebraic 
forms such as dual-quaternions.

Furthermore, the pyGLV package contains the implementation of a set of systems, responsible to digest 
the scenegraph and apply various tasks of the CG pipeline, such 
as the evaluation of the model-to-world and root-to-camera matrices, 
through scenegraph traversals.
\begin{enumerate}
   \item \textbf{TransformSystem:} A system traversing the scenegraph, 
   evaluating the local-to-world matrix for each object-entity. This is 
   accomplished by multiplying all Transform 
   component matrices, starting from a given node-object and 
   ascending the graph to its root-node 
   (see Fig.~\ref{fig:scenegraph}); the 
   order of matrix multiplication is the same to the one taught 
   in the theoretical CG course; the first transformation matrix is the right-most in this product.
   \item \textbf{CameraSystem:} A system evaluating the 
   root-to-camera matrix in a CG scenegraph. 
   This crucial for the CG pipeline matrix 
   is evaluated via this system, by identifying the single node with 
   a camera component and returning the inverse of the model-to-world 
   matrix for that entity (see Fig.~\ref{fig:scenegraph}). 
   \item \textbf{InitGLShaderSystem:} A system used to 
   make data initializations 
  and \emph{start the CG engine}, outside the 
   main rendering loop.
   \item \textbf{RenderGLShaderSystem:} A system responsible for the main 
   GPU rendering. When both a VertexArray and a Shader 
   component are encountered under the same entity, the  system 
  passes the VAO to the GPU and visualizes  it on screen.
\end{enumerate}
Furthermore, it is possible to decorate these systems which will allow 
the digestion of decorated or 
recently introduced components, depending on the aimed application functionality. As an example, one could create a texture component 
along with the respective system, that would apply the texture 
to a specific object/entity.

\subsection{The pyEEL repository}
The python-Explore-Experiment-Learn (pyELL) repository contains several 
python-based jupyter notebooks, that 
act as a knowledge hub, showcasing basic, cross-platform, scientific 
computing and OpenGL-based real-time computer graphics with applications 
to scientific visualization and deep learning. We envision to extend this repository to a portal for both beginner
and veteran programmers, that want to learn the basics or advanced techniques
in various domains/packages, including python, numpy, git, matplotlib, ECS, 
Geometric Algebra, Graph Neural Networks (GNNs), etc.

\section{Teaching CG with Elements at (under)graduate levels}

In this section we present how Elements has already been utilized in several introductory and advanced CG courses of undergraduate/graduate
students.  
The Elements' features are unfolded sequentially, according to the syllabus of the specific CG course, allowing students to gradually apply in practice the entire set of theoretical knowledge they acquire.

\begin{figure}[tbp]
   \centering
   \includegraphics[width=0.47\textwidth]{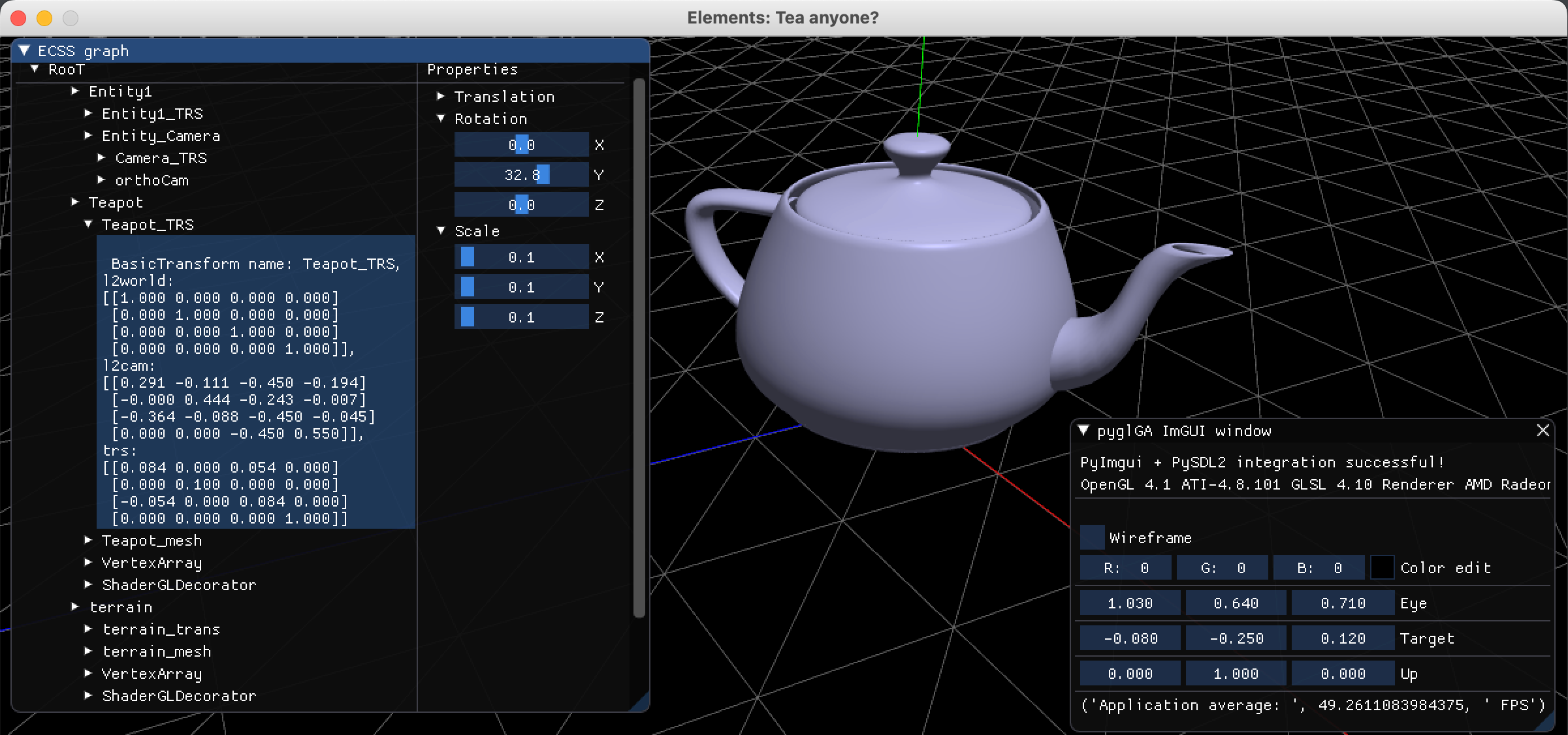}
   \caption{A simple scene with Newell’s teapot being lit by the Blinn-Phong algorithm. The camera (view, target and up) can be changed in real-time via the GUI, generated by the imgui package. 
   An introductory step towards understanding camera, primitives 
   and scene inspection and manipulation. }
   \label{fig:teapot}
\end{figure}

After the introductory lesson to CG and its importance (week 1), students
are taught the basic mathematics involving homogeneous coordinates, rigid body transformations, projections and viewing matrices (weeks 2 \& 3). Since 
enrolled students are at least in their 3rd year, this is simply 
an extension of their knowledge of linear and vector algebra. In week 4, 
they are introduced to graphics programming during the 1st Lab, involving 
the setup of the Elements project and experimenting with basic mathematical 
functions. After learning more on geometry and polygonal modeling, as well 
as scene management (week 5), the 1st assignment is released (see Section~\ref{sec:added_value}) and the 2nd lab takes place, where the initial 
Elements examples are explained. As the course progresses, students 
learn about hardware lighting \& shading, advanced GLSL (week 6), 
textures and basic animation (week 7). The 3rd lab (week 7) then exposes 
students to the respective Elements examples that demonstrate the Blinn-Phong lighting algorithm, texture loading, object-importing and rigid-skinned model animation 
(eventually understanding all principles behind a scene such as  Fig.~\ref{fig:teapot}). The relative 2nd assignment is released on week 8 (see Section~\ref{sec:added_value}). 

The rest of the course unfolds without using Elements, as students must also get acquainted with a MGE such as Unity and even 
more complex frameworks. For completeness, we mention the remaining 
topics of the CG curriculum: Visible surface 
determination and real-time shadows are taught at week 8, followed by ray-tracing 
and CG in 3D games (week 9). The remaining weeks 10-13 involve Unity lectures \& tutorials, a lecture on VR \& AR techniques, along with a hands-on lecture 
with VR \& AR head-mounted displays. A Unity-based assignment is released on week 10.

\section{Elements framework student assignments}

In this section, we describe few key graduate and undergraduate 
assignments that were developed using the Elements project.

\subsection{Skinned Animated Objects in Elements}

\begin{figure}[tbp]
   \centering
   \includegraphics[width=0.47\textwidth]{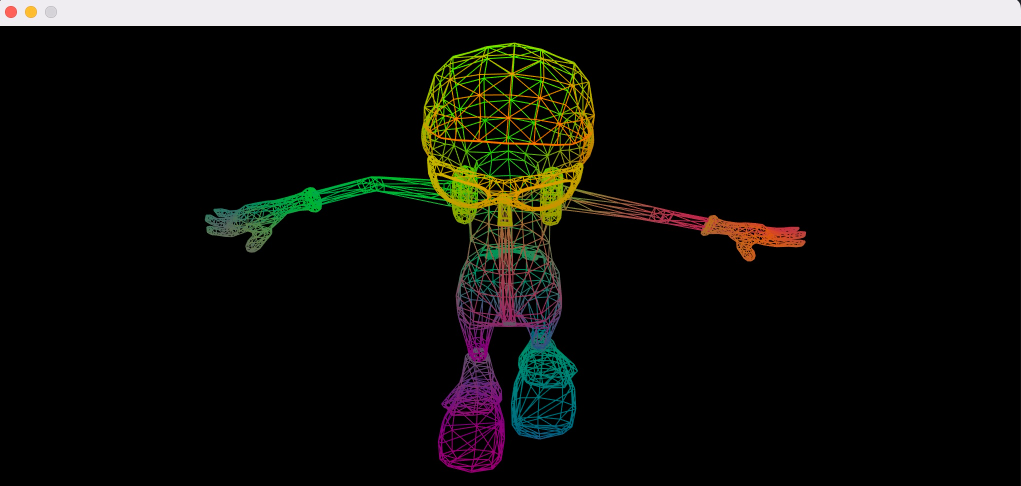}
   \caption{A rigged model imported and animated
   using the animation equation. A dedicated component and system is responsible for storing and digesting the animation model and data.}
   \label{fig:skinned}
\end{figure}

In this assignment, students extended the basic mesh component to 
a SkinnedMesh component and created a respective 
system to allow animation of rigged animated models in Elements. 
The new component allows the storage of skinned 3D model data, 
which are loaded via the \texttt{pyassimp} package. The corresponding 
system  digests such components and applies the animation equation 
to the skinned model between two keyframes (see Fig.~\ref{fig:skinned}). 

\subsection{Geometric Algebra Support for Elements} 
\label{sub:representation_form_agnostic}

This assignment dealt with the creation of a component and a system 
that enables the utilization of Geometric Algebra. Current CG teaching 
frameworks \cite{papagiannakis2014glga} employ 
typical representation forms to describe the translational, scaling  and 
rotational information of each object. This involves vectors or matrices
for the translation/scale matrices, and euler angles or quaternions 
for rotation. Regardless of the form chosen, all data 
must be transmuted to the equivalent matrix, that will be sent to the 
GPU shader. Using the developed component and system, introduced in this work, 
it is possible to use Geometric Algebra (GA) \emph{multivectors} 
to represent such transformations (translations, rotations and 
dilations), which are an evolution step of current representation forms \cite{LessIsMore}. As such, the use of GA-based forms (motors, rotors and dilators) as 
decorated Transform components, and their digestion from the decorated Transform system, can be accomplished with no modifications in the existing pipeline or code that renders 
the scene. In fact, the programmer may use diverse algebraic
formats (matrices, vectors, quaternions, dual-quaternions, 
multivectors) to define the Transform or Mesh components of an object, 
with minimal programming effort, since systems digesting such components work independently and in parallel (see Fig.~\ref{fig:scenegraph}). 

We use geometric algebra in order to teach our students a unifying model for transformations, instead of using 3 algebras: euclidean algebra, quaternion algebra, dual-quaternions in affine geometry. 
Especially referring to Projective and Conformal GA, multivectors can be used to express both translations, rotations as well as 
dilations (uniform scalings, exclusive to CGA) and are suitable for 
CG rendering and animation \cite{papagiannakis2013geometric} .


\subsection{Graph Neural Network for Elements} 
\label{sub:graph_neural_network}

In the context of a graduate CG course, students were assigned to 
investigate a connection between Elements and trending machine/deep 
learning techniques. Google's SceneGraphFusion \cite{Wu2021SceneGraphFusionI3}, which extracts semantic scene graphs from a 3D environment generated from RGB-D images, highlight the importance of graph representations of 3D scenes in the process of labeling and extracting hierarchical data from the graph. 
As Elements is based on a scenegraph representation, the identification of equivalent graphs, that  
store similar information, and their subsequent feeding 
to a suitable GNN, would lead the way to 
performing a simple predictive task. 

Manipulating the Elements scenegraph, students were able to 
explore various equivalent graph representations, that could be used as input to 
graph-based machine/deep learning models. A major task in this process was to acquire a clear understanding of node, 
edge and graph features and how these could be derived from the original 
entities and components of the scenegraph. As an example, mesh components stored in the scenegraph, holding  vertex positions and the face indices, can be transmuted to subgraphs, where nodes correspond to vertices, and node features to edges extracted through the face indices (neighbouring vertices correspond to neighbouring nodes). 
An alternative approach would store an entity's location with respect to its parent-entity as an edge feature, holding 
the corresponding translation, rotation and scale vertex, 
and replace the mesh data by a label (such as ``cube'' or ``pyramid''), that could either be given or predicted by another GNN. 

\begin{figure}[btp]
   \centering
   \includegraphics[width=0.45\textwidth]{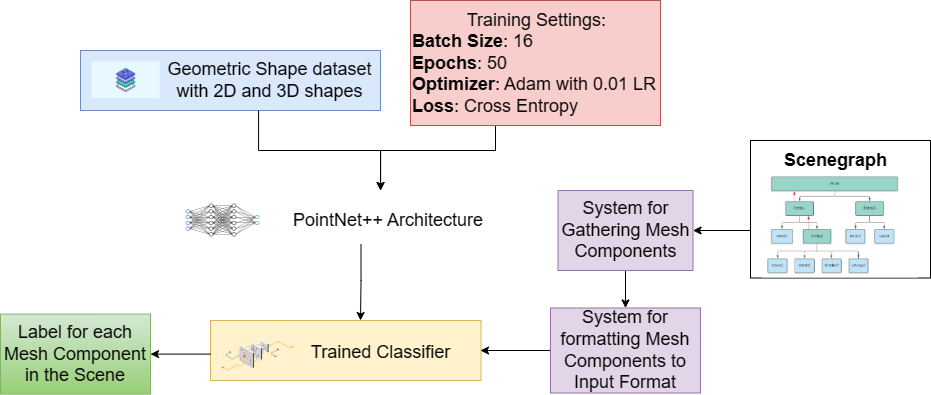}
   \caption{GNN training process - Object labelling using ECSS.}
   \label{fig:ElementsGNN}
\end{figure}

Depending on the different graph representation form, various tasks 
were performed using standard GNN techniques. For example, in the first 
representation described above, a GNN could successfully identify 
if a specific shape, for example ``a cube'', existed in the scene, whereas, 
the second representation form allowed the prediction of spatial 
relationships such as ``is there a pyramid on top of a cube?''. 
Since 3D models can also be seen as point clouds, after a certain reformatting pre-processing step,  we fed them to a classifier, based on the PointNet++ \cite{qi2017pointnet++} architecture, for point cloud classification.  The model was trained for 50 epochs on a dataset of 40 different types of Geometric objects (2D and 3D) and achieved over 90 percent of accuracy on the test set. After completion of the training process (see Fig. \ref{fig:ElementsGNN}), the model could easily be integrated in any Elements project, to classify the different GameObjects in the scene.

As an ongoing graduate project, a generative artificial intelligence 
approach is being investigated. By using suitable representations and 
employing more complicated ML/DL techniques (involving transformers, 
graph convolutional networks, etc.) we aim to provide a model that 
can generate a complicated scenegraph based on user text or voice input, 
e.g., ``a ring of cubes'', ``a solar system with two centers'', 
``the interior of a house'' or ``a virtual operating room''. The auto-creation of 
such scenes is not straightforward, and should be decomposed in a set of 
simpler tasks, such as the generation of 3D objects and their spatial 
arrangement.

\section{Using Elements in existing CG curricula} 
\label{sec:added_value}

The Elements project has been employed as a teaching material for 
a graduate (CS-553 Interactive Computer Graphics) and an undergraduate (CS-358 Computer Graphics) CG course at the University of Crete (UoC), Greece 
as well as for the undergraduate (ECE-E34 Computer Graphics) CG course 
at the University of Western Macedonia (UoWM), Greece. The number of students 
actively participated in these assignments, using the Elements project, 
were 28, 24 and 24 respectively.

\begin{figure}[btp]
   \centering
   \includegraphics[width=0.45\textwidth]{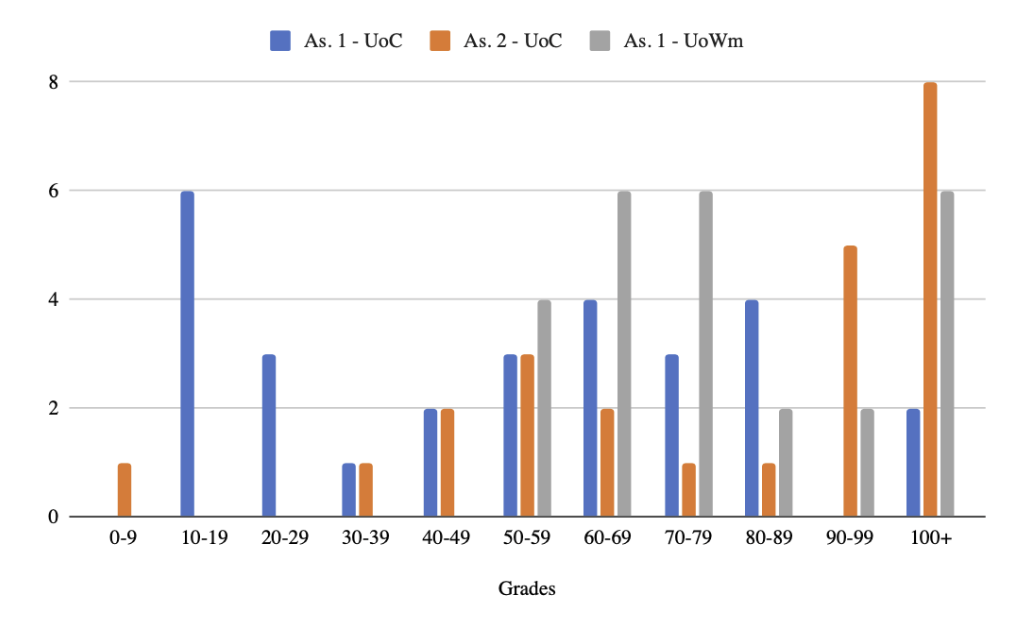}
   \caption{The student grades of the two UoC and one UoWM Elements-based assignments. The horizontal axis shows the grade ranges and the vertical the number of students within each range.}
   \label{fig:grades}
\end{figure}

There were two assignments within the undergraduate course taught in 
UoC that regarded the use of the Elements project. 
The first one, given on week 5, regarded simple tasks such as using the built-in functions for transformations (translate, rotate and dilate),
for camera-related functions (\texttt{lookat} and \texttt{perspective}), as well as generation of models (polygons and sphere) in a constructive way.
The second one was given on week 8 and involved evaluation of normals via model processing, scene manipulation via a GUI system and implementation of the Blinn-Phong model.
Although there were 4 less participants in the second assignment (28 in 
the first and 24 in the second), the grades indicate that students  
quickly gain expertise on the Elements project, just a few weeks after, and managed to perform better, even in more demanding subjects  (see Fig.~\ref{fig:grades}, comparing orange and blue bars).

In UoWM, the traditional approach of teaching OpenGL in C/C++, without the
use of a specific framework, was augmented with Elements. Students, having
already acquired knowledge on basic OpenGL programming, attended three 2-hour
laboratory lectures in order to get acquainted with Elements and were 
given a single, optional, related assignment, aiming to evaluate and compare students' understanding on basic and advanced CG pipeline stages. The 
assignment involved the built a 3D scenegraph, along with the generation of custom polygonal 3D models, a GUI system that allows transformations and animations applied on a single or group of 3D objects, and the Phong illumination model through GPU scripting. Even after such a short introduction, the 24 students that participated managed to easily adapt. Course evaluations highlighted the direct relation of Elements with the taught theory, compared to the traditional OpenGL teaching approach; their grades, shown as gray bars in Fig.~\ref{fig:grades}, are quite above average.

\section{Performance} 
\label{sec:performance_and_comparison_to_unity}

Using Elements, the rendering of moderately complex scenes is possible in 
real-time; in our tests, we managed to load up to 150 objects, each one of approximately 2900 vertices, without dropping below 60fps (tested with a 
MacBook Pro, 2.6 GHz 6-Core Intel Core i7, 16 GB 2667 MHz DDR4, Intel UHD Graphics 630 1536 MB). All parts of the implementation are intentionally written in python, to promote student understanding and allow the fast 
 creation of proof-of-concept examples. 
Our primary goal when designing Elements was the maximization 
of the educational impact rather than the rendering performance.


\section{Conclusions and Future Work}

In this work, we have presented Elements, a novel open-source pythonic entity-component-systems in a scengraph framework, suitable for scientific, visual as well as neural computing. 
The python packages comprising Elements: pyECSS, pyGLV and pyEEL include 
the basic implementation of such an approach along with useful 
examples that may be used to quickly introduce even inexperienced CG programmers
to main CG-programming principles and approaches. Despite its simplicity, 
it remains white-box, allowing users to dive in the graphics pipeline and
tamper with any CG pipeline step. Empowered by python's rapid prototyping and 
developing ability, users may implement new or decorated components and/or systems to 
extend the functionalities offered by Elements. 
The impact of Elements' features (current and future) 
on various scientific domains/packages are demonstrated
in jupyter notebooks, in the pyEEL repository, towards creating a 
learning hub for both novice and intermediate developers.

Taking advantage of the ECSS underlying system, this 
project aims to bridge CG curricula and MGEs, that are based on the same approach. This will prepare 
students, that develop their CG knowledge, to take the next step to game engines. The bridging will also enable rapid prototyping in Elements and a subsequent easy
port to game engines, to further evaluate its performance 
or other metrics, not accessible in a python environment.

Work in progress involves the integration of the Elements project 
with open source packages, such as OpenXR, that will allow the 
3D scene to be output in virtual or augmented reality head-mounted 
displays, which will in turn allow enhanced data visualization and, 
ultimately, immersive analytics. Furthermore, the 
connection with modern Unity-based frameworks such as 
\cite{MAGES4} are also actively explored. 
As major neural deep learning frameworks are python-based,
we aim to provide more powerful and versatile ways of 
scientific visualization (input/output, to/from). Furthermore, 
we will strengthen the connection of Elements with GNN-based 
geometric deep learning techniques towards performing more 
complex tasks, such as generating a new ECSS.

A subjective grade-based comparison with previous years is not feasible, 
as different frameworks, assignments and syllabus were used back then.
However, we will keep on monitoring the benefits that Elements can 
bring to CG curricula that adopt its use and conduct further 
evaluations on student reactions as well as randomized control trials,
towards highlighting the added pedagogical values of using the proposed project 
instead of classical C/C++ or other frameworks.

\section*{Acknowledgements}
We would like to thank Irene Patsoura and Mike Kentros for their contribution, related to the assignments presented in 
this paper.

\section*{Code Availability}
Elements is available via open-source at 
\url{https://papagiannakis.github.io/Elements/}.

\balance
\bibliographystyle{plain} 
\bibliography{bibliography}       

\end{document}